# TiO$_2$ nanotubes with laterally spaced ordering enable optimized hierarchical structures with significantly enhanced photocatalytic H$_2$ generation

Nhat Truong Nguyen,[a,†] Selda Ozkan,[a,†] Imgon Hwang,[a] Anca Mazare,[a] Patrik Schmuki[a,b]*

[a]Department of Materials Science and Engineering WW4-LKO, University of Erlangen-Nuremberg, Martensstrasse 7, D-91058 Erlangen, Germany.
[b]Chemistry Department, Faculty of Sciences, King Abdulaziz University, 80203 Jeddah, Saudi Arabia Kingdom
[†] These authors contributed equally to this work
*Corresponding author. Email: schmuki@ww.uni-erlangen.de

**Abstract**

In the present work we grow self-organized $TiO_2$ nanotube arrays with a defined and controlled regular *spacing* between individual nanotubes. These defined intertube gaps allow to build up hierarchical 1D-branched structures, conformally coated on the nanotube walls using a layer by layer nanoparticle $TiO_2$ decoration of the individual tubes, i.e. having not only a high control over the $TiO_2$ nanotube host structure but also on the harvesting layers. After optimizing the intertube spacing, we build host-guest arrays that show a drastically enhanced performance in photocatalytic $H_2$ generation, compared to any arrangement of conventional $TiO_2$ nanotubes or conventional $TiO_2$ nanoparticle layers. We show this beneficial effect to be due to a combination of increased large surface area (mainly provided by the nanoparticle layers) with a fast transport of the harvested charge within the passivated 1D nanotubes. We anticipate that this type of hierarchical structures based on $TiO_2$ nanotubes with adjustable spacing will find even wider application, as they provide an unprecedented controllable combination of surface area and carrier transport.

Since the groundbreaking work of Fujishima and Honda[1] in 1972, photocatalytic $H_2$ generation by water splitting based on $TiO_2$ materials has received immense research interest. Over the past decades, not only $TiO_2$ but also hundreds of other semiconductors have been investigated for their $H_2$ generation performance (e.g. ranging from Si to group II-VI, III-V compound semiconductors and to various transition metal oxides, and so on).[2–4] In spite of some intrinsic deficits of $TiO_2$, such as a wide band gap and a sluggish electron transfer kinetics, it still remains the most studied photocatalytic material as it has an almost unique (photo)corrosion resistance and comparably low cost.[5–9] The photoelectrochemical principle to hydrogen generation is that suitable light irradiation promotes electrons from the semiconductor ($TiO_2$) valence band to the conduction band, thus generating electron hole pairs that then can react at the semiconductor surface with redox couples in the surroundings – hydrogen ($H_2$) generation occurs from ejected conduction band electrons to $H_2O$ or $H^+$.

In order to maximize the efficiency of the $H_2$ photocatalytic reactions on $TiO_2$, typically $TiO_2$ nanoparticles, either free floating in the solution or compacted to electrodes, are used.[10–13] Nanoscale photocatalysts provide not only a high specific surface area but also a short diffusion path for the excited carriers to reach the particle surface (minimizing recombination). More recently, one-dimensional (1D) structures such as nanotubes, nanowires and nanorods have attracted significant interest as they can provide, except for a large surface area, a *directional* carrier transporter (faster), and in many cases intrinsically support an orthogonal electron-hole separation.[5,7,8,14] These features further aid in minimizing the recombination of charge carriers and thus minimize the efficiency loss.[5,8,9]

In the past years, particularly highly ordered $TiO_2$ nanotube layers (NTs) grown by self-organizing electrochemical anodization of titanium have attracted wide interest.[8] For these tubes, which are vertically aligned on the substrate, geometry and structure can be easily controlled by the anodization parameters and by a post formation thermal annealing.[15] In some cases, such nanotube layers were reported to outperform comparable nanoparticle layers in their photocatalytic performance.[16,17] This was mainly ascribed to a more efficient charge separation and transport which, in fact, is not surprising considering that the electron mobility of nanoparticle layers has been reported to be orders of magnitude lower than that of nanotubes.[16,18–20] Nevertheless, in terms of specific surface area, such NTs provide clearly lower values (30-50 $m^2\ g^{-1}$) than comparable nanoparticle layers where values of 100 $m^2\ g^{-1}$ (and more) can be reached.

In order to combine the benefits of NTs (electron transport properties) and nanoparticles (large surface area), hierarchical structures have been explored where $TiO_2$ particles are decorated onto the $TiO_2$ NTs.[21] However, up to now, optimized hierarchical structures require the ability to space the 1D conducting material in desired distances on a substrate, e.g. as in the case of nanowires,[22] nanorods[23] or nanotubes grown by template method.[24] However, using classic anodic $TiO_2$ nanotube layers does not allow this, as conventional anodization leads to a hexagonally close-packed arrangement of nanotubes, and this geometry allows only a limited decoration with defined secondary nanoparticle layers.

In the present work we introduce the use of *spaced*, self-ordered nanotube layers with controllable regular intertube gaps to build defined hierarchical structures, as shown in Fig. 1. In contrast to the classic (hexagonal close-packed) NTs in Fig. 1c, spaced nanotubes can be precisely coated, layer by layer with nanoparticles, up to an optimized number of layers without clogging the tube's mouth. These structures when used for photocatalytic $H_2$ production can exhibit a strongly enhanced efficiency compared with any conventional $TiO_2$ NTs or compared to a plain $TiO_2$ nanoparticle layer.

The formation of self-organized nanotubular layers with defined intertube gaps takes place in some electrolytes under anodization conditions (outlined in the SI) that establish a controlled spacing, in fact, independent control of tube diameter and intertube distance, can be established by control over the water content in the electrolyte and the applied voltage. After a set of preliminary experiments (see SI) where we evaluated the photocurrent response of nanotubes that have different spacing, we selected nanotubes with spacing in the range of 150-200 nm in order to build up hierarchical structures (see Fig. S1, S2).

The spaced TiO$_2$ NTs were grown by anodization of Ti in hot triethylene glycol (TEG) based electrolyte, while the reference nanotubes (R-NTs) are grown in ethylene glycol based electrolyte, see Fig. 1 (for more details, see SI). After anodization, these spaced TiO$_2$ NTs (S-NTs) uniformly cover the surface (Fig. 1a), and have a 220 nm diameter with an average intertube space (wall-to-wall) of approx. 150-200 nm (Fig. 1b). Such a geometry was found to be ideal for a highly conformal layer by layer decoration with TiO$_2$ nanoparticles (NPs), using a TiCl$_4$-hydrolysis approach,[25] as shown in Fig. 1e and Fig. 2: it allows a controlled decoration with up to 5 layers. Each decorated layer is ≈17 nm thick and consists of ≈5-10 nm diameter TiO$_2$ nanoparticles that homogeneously coat both the interior and exterior surfaces of the nanotube walls, as evident from the SEM images in Fig. 1d and e. After annealing the decorated NTs at 450°C in air, high resolution TEM images (Fig. 1f,g) show that the TiO$_2$ NPs are interconnected and fully crystallized (anatase (101) crystallographic planes, 3.46 Å lattice spacing) can be identified. Moreover, XRD confirms the anatase phase of the NTs (see Fig. 2.c), while XRD and TEM results together indicate that not only the tubes but also the decorated particles were converted to anatase.[26,27]

In order to assess the photocatalytic H$_2$ evolution performance, layers were decorated with a Pt co-catalyst (1 nm Pt thick, SI) and illuminated with continuous UV light (325 nm) in an ethanol-water mixture (Fig. 1h,i). Fig. 1h shows a comparison of the H$_2$ amounts evolved from spaced NTs with and without nanoparticle decoration, reference NTs and a nanoparticle (P) film. The H$_2$ evolution amount from plain S-NTs is already slightly higher than R-NTs, however both are clearly lower than for the defined TiO$_2$ nanoparticle film on FTO substrate (the nanoparticle film is shown in Figure S4).

The detailed evaluation of the spaced NTs hierarchical structures, with layer by layer TiO$_2$ nanoparticle decoration, is shown in the SEM images of Fig. 2a. With every additional layer, not only the inner diameter of NTs decreases but also the outer diameter expands. More importantly, the photocatalytic performance increases and maximizes after three layers, then decreases (Fig. 2b); higher amounts of NP loadings led to a drop in $H_2$ generation as NTs open geometry decreases, thus reducing the surface area as well as light penetration.

However, for S-NTs, after the optimal three times layer by layer $TiO_2$ NP decoration, a seven times improvement of the photocatalytic $H_2$ generation is obtained and this hierarchical structure clearly outperforms the nanoparticle layer. If the same multiple decoration treatment is attempted with classic NTs (R-NTs), not only does the second decoration already result in an inhomogeneous morphology but also the nanotubes are blocked with $TiO_2$ NPs (Fig. S5). For these R-NTs, the generated $H_2$ amount is ≈16 $\mu mol\ h^{-1}\ cm^{-2}$, which is half of the hierarchical S-NT structure produced under similar conditions.

When investigating the effect of the length of nanotubes for the three times decorated spaced NTs, a maximum $H_2$ production rate is registered for a 7 μm tube length (Fig. 1i, Fig. S6). This is in line with literature, where a thickness of 6-7 μm is found to be an optimum between light absorption and electron diffusion length (previous reports show for NP UV light penetration (325 nm) of 1-3 μm, and diffusion length in $TiO_2$ NTs of several 10 μm while in particle layers this is only some few μm).[20,28,29] Additionally, a 1nm thick Pt decoration is optimal in view of photocatalytic $H_2$ generation (see Fig. S7, S8). For these optimized S-NTs, the $H_2$ generation is linear over extended time, indicating that the structures are stable, see Fig. S9 (moreover, SEM images after extended $H_2$ generation times did not show any significant difference – data not shown).

Usually, as-formed S-NTs are amorphous (only titanium peaks are detected) and by 450 °C air annealing, crystallization induces anatase formation (XRD patterns in Fig 2.c).[30] S-NTs were layer by layer coated with $TiO_2$ nanoparticles and air annealed again at 450 °C (for 10 min) to crystallize the nanoparticles and the hierarchical structures crystallography remains unchanged after the three layers decoration.

In order to characterize the main difference between the hierarchical structure and conventional tubes, several aspects were investigated which include their photocurrent spectra (Fig. 3a), their relative active surface area (Fig. 3b), the relative electron transport time from intensity modulated photocurrent spectroscopy (IMPS) measurements (Fig. 3c) and their relative ability to photocatalytically generate OH radicals (Fig. 3d).

From photocurrent spectra measurements (Fig. 3a), most apparent is that the nanoparticle layer on FTO shows a very low photon to current conversion efficiency compared with the nanotube layers – this is indicative of the high recombination rate of electron-hole pairs in the nanoparticle structure. Clearly, S-NTs show a significantly higher photocurrent magnitude than R-NTs, with the highest values for the three times loading, and nanoparticle decorated S-NTs show also a shift of the wavelength of the maximum IPCE.

In order to determine the surface area of the structures, dye loading measurements (Fig. 3b) for R-NTs, $TiO_2$ nanoparticle films, spaced NTs and three times loading spaced NTs (S-NTs-T3) were performed and resulted in 41, 101, 24 and 131 nmol cm$^{-2}$, respectively. This shows that the NPs decoration of S-NTs strongly increases their specific surface area, up to 5 times (from dye loadings of 24 to 131 nmol cm$^{-2}$), thus enhancing the active surface area. In fact, the average particle size in the nanoparticle layers (10-15 nm) is larger than the particles obtained from the $TiCl_4$ layers (5-10 nm), therefore the hierarchical structure shows a higher surface area than the nanoparticle layer.[31]

To examine the electronic properties of the structures, IMPS measurements were performed for these selected structures, see Fig. 3c. Clearly, electron transport in the $TiO_2$ particle layer is significantly slower than that for either bare nanotubes – reference or spaced. While both types of non-decorated nanotubes (hexagonally-packed and spaced) have a similar electron transport rates, one layer NP decoration of the spaced NTs improves the electron transport time by a decade and the optimal three times decoration shows only slightly faster transport time compared to one layer, while additional NP layers deposition leads to no further improvement. These findings can be ascribed to the fact that the $TiCl_4$ treatment not only increases the specific surface area but also passivates the defects on $TiO_2$ NTs, thus enhancing the electronic properties.[32] A classic photocatalytic test for the formation of OH• radicals, i.e. using fluorescence from terephthalic acid under UV light ($\lambda$=325 nm) illumination – see Fig. 3d, confirms that under the same conditions in spaced NTs (S-NTs) twice the rate of OH• radicals can be formed than with classical hexagonally packed NTs (R-NTs) – this confirms that not only a strongly improved $H_2$ formation rate can be achieved using hierarchical NTs but also that in classic photocatalytic reactions (photodegradation, photosynthesis) hierarchical structures are able to generate strongly enhanced amounts of active species.

In summary, the present work shows the fabrication of a photocatalytic platform consisting of hierarchical $TiO_2$ nanostructures. The approach is based on using highly ordered spaced $TiO_2$ NTs as a scaffold for a controlled layer by layer $TiO_2$ nanoparticle deposition. After optimizing the nanoparticle decoration, hierarchical $TiO_2$ nanostructures show a significant enhancement of the photocatalytic performance in comparison to the hexagonally close-packed $TiO_2$ NTs and conventional $TiO_2$ nanoparticle films. This improvement is ascribed to the combination of fast electron transfer due to the 1D structure and to the high surface area owing to nanoparticle decoration. This also opens a practical route for the deposition of other materials in the spaced $TiO_2$ NTs, thus using to full advantage such morphology for the development of other state-of-the art applications.

**Acknowledgements**

Dr. Lei Wang is acknowledged for helping in the evaluation of the diffraction data. Xuemei Zhou is acknowledged for XPS measurements. Prof. Kiyoung Lee and Jeong Eun Yoo are acknowledged for valuable discussions. The authors would also like to acknowledge the ERC, the DFG, the Erlangen DFG cluster of excellence EAM, and the DFG funCOS for financial support.

32.	S. So, I. Hwang and P. Schmuki, *Energy Environ. Sci.*, 2015, **8**, 849–854.

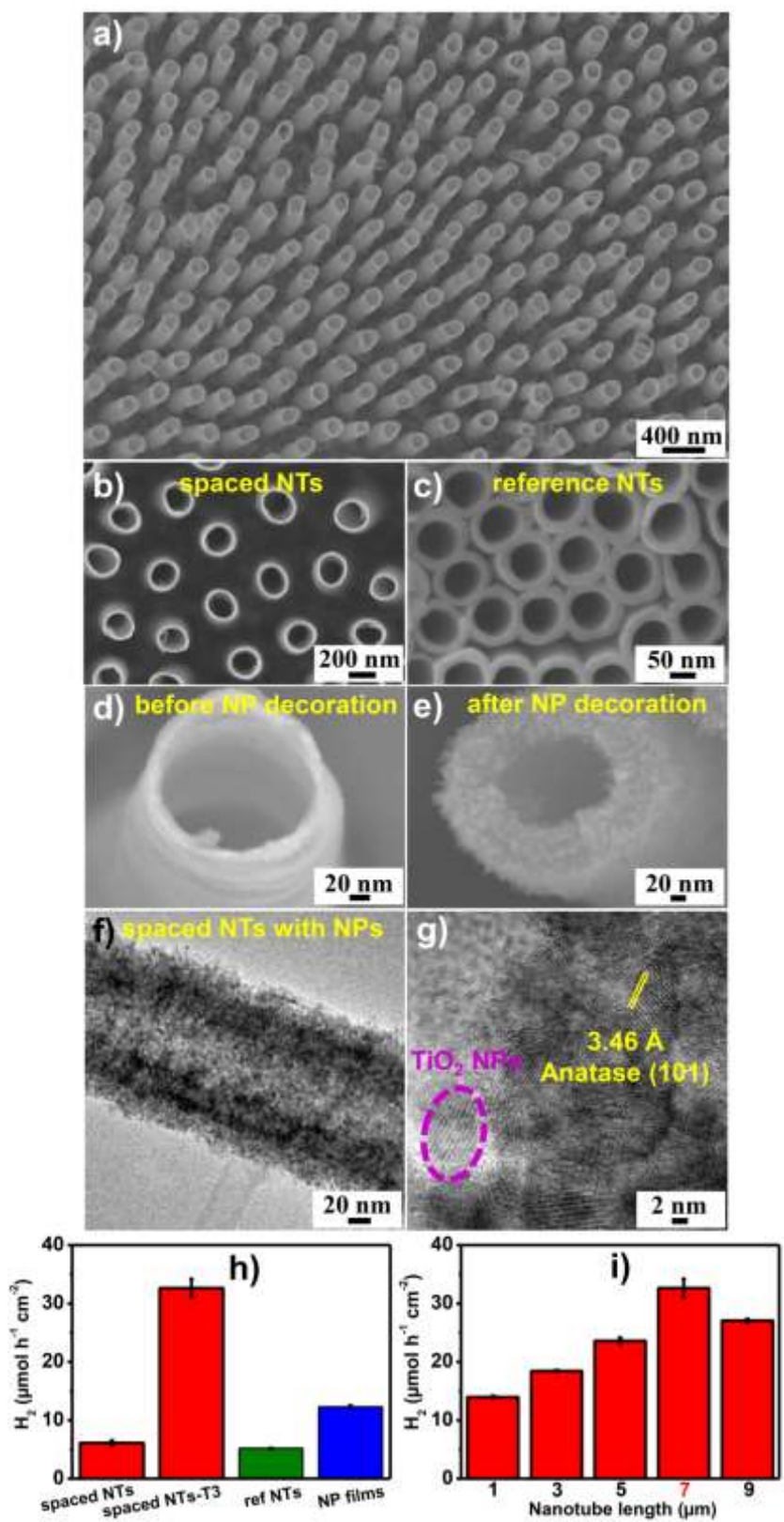

**Figure 1.** SEM images of: a), b) spaced $TiO_2$ NTs; c) reference $TiO_2$ NTs; d) spaced $TiO_2$ NTs before and e) after 3 layers of $TiO_2$ nanoparticle decoration. f) and g) TEM images of spaced $TiO_2$ NTs with $TiO_2$ NPs. Photocatalytic $H_2$ evolution measured for: h) spaced $TiO_2$ NTs with/without $TiO_2$ nanoparticles, reference $TiO_2$ NTs and $TiO_2$ nanoparticles on FTO; all samples are decorated with 1 nm-thick Pt layers; i) spaced $TiO_2$ NTs with different thicknesses after decoration with 3 layers of $TiO_2$ nanoparticles and 1 nm-thick Pt layers.

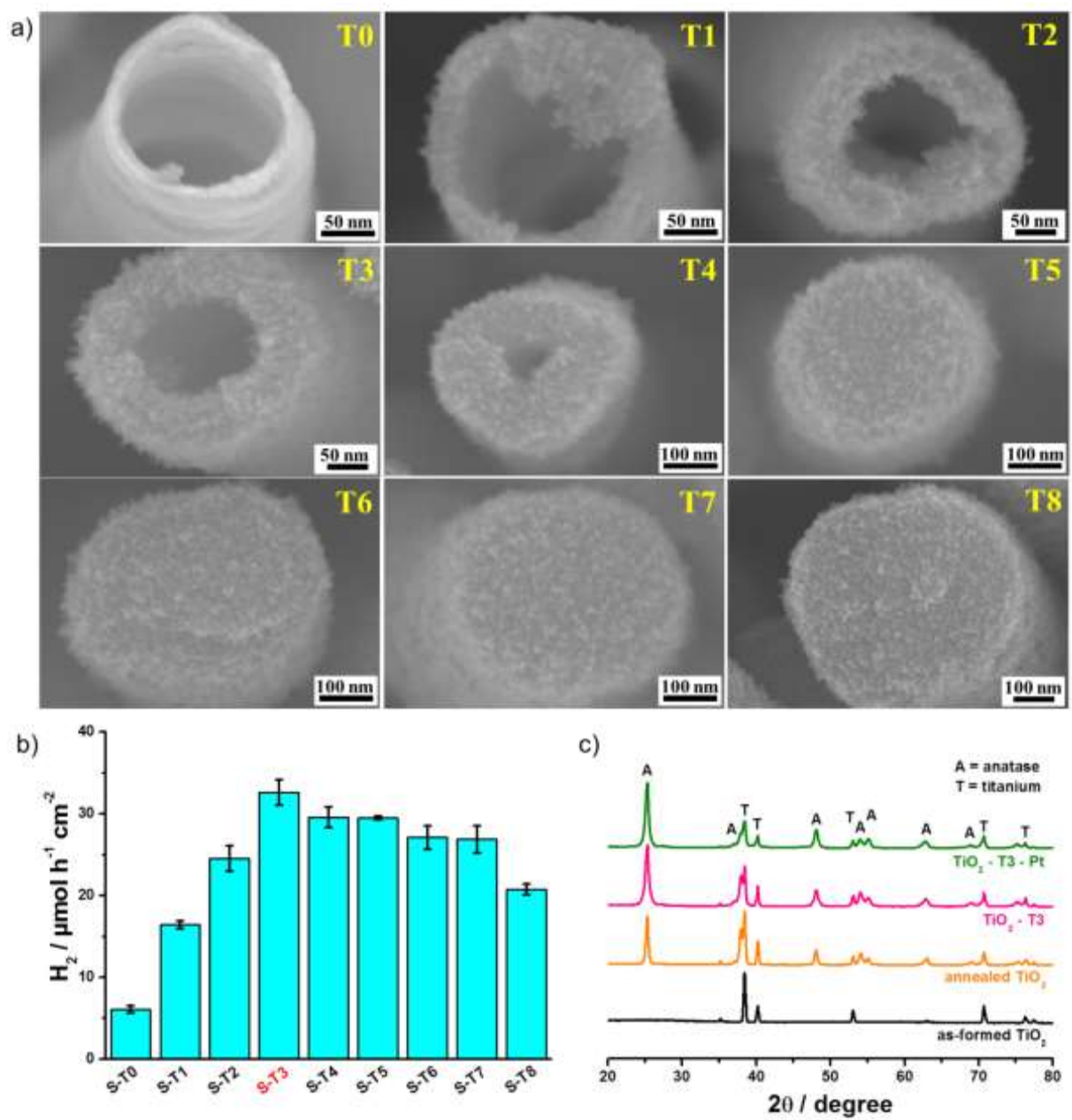

**Figure 2.** a) SEM images and b) photocatalytic $H_2$ evolution measured for spaced $TiO_2$ NTs decorated with different layers of $TiO_2$ nanopartciles (all samples were decorated with 1 nm-thick Pt layer); c) XRD patterns of as-formed $TiO_2$, annealed $TiO_2$, $TiO_2$ NTs decorated with $TiO_2$ nanoparticles ($TiO_2$ – T3), and $TiO_2$ NTs decorated with $TiO_2$ nanoparticles and 10 nm-thick Pt layer ($TiO_2$ – T3 – Pt).

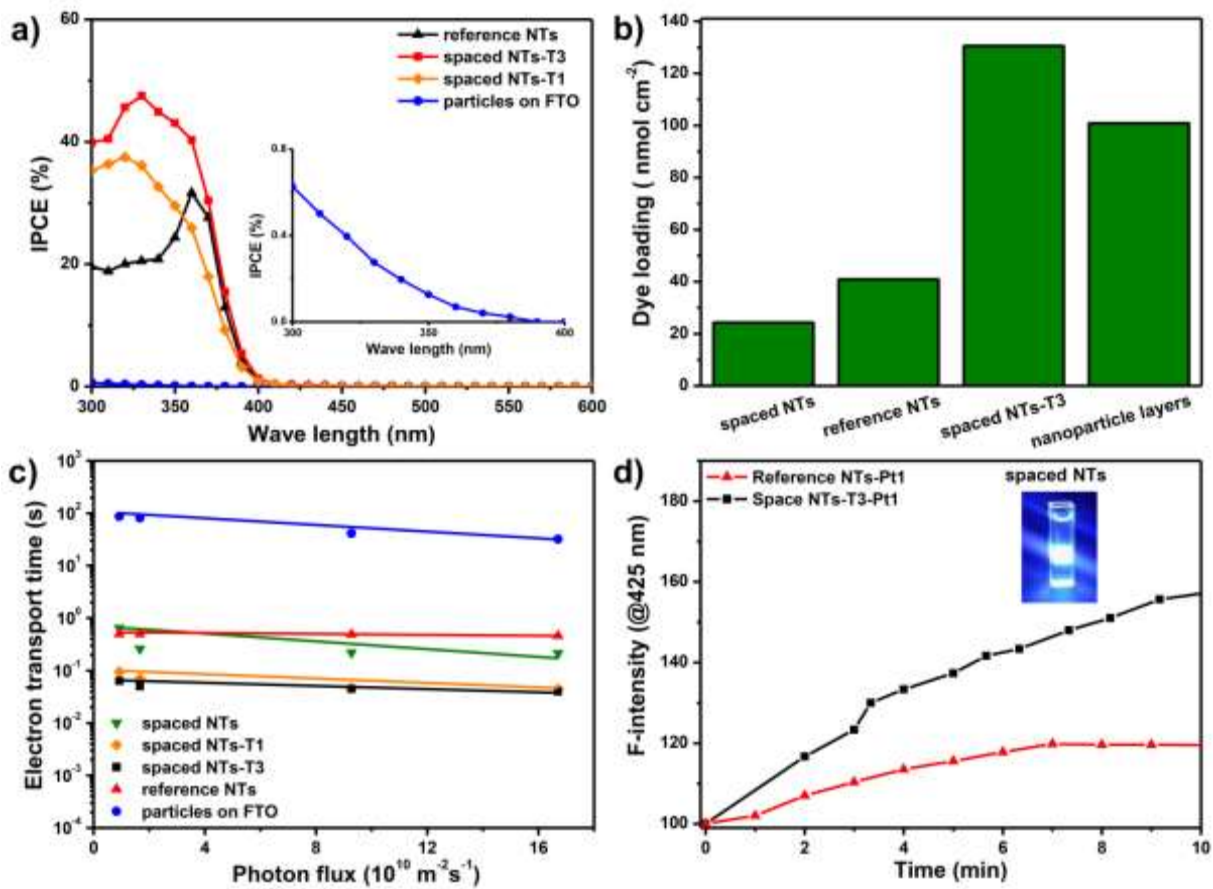

**Figure 3.** a) IPCE spectra in 0.1 M $Na_2SO_4$ at 0.5 V (vs. Ag/AgCl): one time and three times decorated spaced NTs, reference NTs and $TiO_2$ particles (prepared by doctor blading method); b) dye loading measurements of a) samples; c) electron transfer time constants from IMPS measurements under UV light illumination (369 nm); d) Fluorescence intensity (measured at 425 nm) changing with time observed during UV illumination (excitation at 325 nm) in terephthalic acid solution. The inset shows optical image of fluorescence test for spaced NTs.

# TiO$_2$ nanotubes with laterally spaced ordering enable optimized hierarchical structures with significantly enhanced photocatalytic H$_2$ generation

Nhat Truong Nguyen,[a,†] Selda Ozkan,[a,†] Imgon Hwang,[a] Anca Mazare,[a] Patrik Schmuki[a,b]*

[a]Department of Materials Science and Engineering WW4-LKO, University of Erlangen-Nuremberg, Martensstrasse 7, D-91058 Erlangen, Germany.
[b]Chemistry Department, Faculty of Sciences, King Abdulaziz University, 80203 Jeddah, Saudi Arabia Kingdom
[†] These authors contributed equally to this work
*Corresponding author. Email: schmuki@ww.uni-erlangen.de

**Experimental section**

*Growth of TiO$_2$ nanotubes*: Titanium foils (Advent Research Materials, 0.125 mm thickness, 99.6+% purity) were degreased by sonication in acetone, ethanol and deionized water, followed by drying in N$_2$ gas stream. The TiO$_2$ nanotubes were formed by anodizing titanium foils in triethylene glycol electrolyte containing 0.3 M NH$_4$F and 3 M H$_2$O at 60 V for 2 h at 60 °C. The DC potential was applied by using a VLP 2403 pro, Voltcraft power supply. After the anodization, the samples were soaked in ethanol for few hours, and then dried under N$_2$ gas stream. Subsequently, the TiO$_2$ nanotubes were annealed at 450 °C in air for 1 h using a Rapid Thermal Annealer (Jipelec Jetfirst 100 RTA), with a heating and cooling rate of 30 °C min$^{-1}$. For reference TiO$_2$ NTs, titanium foils were anodized in ethylene glycol electrolyte containing 0.15 M NH$_4$F and 3 wt% H$_2$O, at 60 V for 15 min. The R-NTs were then annealed in air at 150°C for 1 h, followed by piranha treatment at 75°C for 2.5 min to increase inner diameter of the NTs. All piranha-treated samples were annealed in air at 450°C for 1 h.

*TiO$_2$ nanoparticle decoration*: For TiCl$_4$ treatments, aqueous solution of 0.1 M TiCl$_4$ was prepared under ice-cooled conditions. The TiO$_2$ nanotube layers were then treated in a closed vessel containing of 7 mL TiCl$_4$ solution at 70 °C for 30 min. Subsequently, the samples were washed with distilled water and rinsed with ethanol to remove any excess TiCl$_4$, and finally dried in a N$_2$ stream; and this process was replicated many times. After this treatment, the decorated samples were annealed again at 450 °C for 10 min (Rapid Thermal Annealer) to crystallize the attached nanoparticles.

*TiO$_2$ nanoparticle layers*: TiO$_2$ nanoparticle layers (Ti-nanoxide HT, Solaronix) were prepared by doctor blading method on fluorine-doped tin oxide glass (FTO).

*Pt nanoparticle decoration*: In order to decorate the TiO$_2$ NTs, plasma sputter deposition (EM SCD500, Leica) was used to deposit different amounts of Pt. The amount of Pt was controlled by measuring their nominal thickness with an automated quartz crystal film-thickness monitor.

*Characterization of the structure*: Field-emission scanning electron microscope (FE-SEM, Hitachi S4800) was used to characterize the morphology of the samples. The chemical composition of the samples was analyzed by X-ray photoelectron spectroscopy (XPS, PHI 5600, US). X-ray diffraction (XRD) performed with a X′pert Philips MPD (equipped with a Panalytical X'celerator detector) using graphite monochromized Cu K$_\alpha$ radiation ($\lambda = 1.54056$ Å), was employed to examine the crystallographic properties of the materials. The fluorescence intensity was measured by a using a Newport Optical Power Meter at 425 nm. The hydroxyl radicals (•OH) detection was conducted by illumination of UV light (325 nm) on the samples immersed in aqueous solution of 3 mM terephthalic acid, 0.01 M NaOH and 0.1 M NaCl.

*Photocatalytic measurements*: Photocatalytic measurements were carried out by irradiating the TiO$_2$ nanotube films with UV light (HeCd laser, Kimmon, Japan; $\lambda = 325$ nm, expanded beam size = 0.785 cm$^2$, nominal power of 23 mW cm$^{-2}$) in a 20 vol% ethanol-water solution (for 2 h) in a quartz tube. The amount of generated H$_2$ (which accumulated over time within the quartz tube) was measured by using a gas chromatograph (GCMS-QO2010SE, Shimadzu) equipped with a thermal conductivity detector and a Restek micropacked Shin Carbon ST column (2 m x 0.53 mm). Before the photocatalytic experiments, the reactor and the water-ethanol mixtures were purged with $N_2$ gas for 30 min to remove $O_2$.

*Photoelectrochemical measurements*: Photocurrent spectra were conducted in 0.1 M $Na_2SO_4$ under an applied potential of 0.5 V (vs. Ag/AgCl) in a three-electrode system using 150 W Xe-lamp (Oriel 6365) equipped with a Oriel Cornerstone 7400 1/8 m monochromator (illuminated area=0.785 $cm^2$).

*IMPS measurements*: Intensity modulated photocurrent spectroscopy (IMPS) measurements were conducted using modulated light (10% modulation depth) from a high power LED ($\lambda$ = 325 nm). The modulation frequency was managed by a frequency response analyzer (FRA, Zahner IM6). The photocurrent of the cell was determined using an electrochemical interface (Zahner IM6), and fed back into FRA for analysis.

*Dye loading measurements*: Dye adsorption was conducted by immersing the samples in a 300 mM solution of Ru-based dye (cis-bis (isothiocyanato) bis(2,2-bipyridyl 4,4-dicarboxylato) ruthenium(II) bistetrabutylam-monium) at 40°C for 1 day. The dye solution was a mixture of tert-butyl alcohol and acetonitrile. Subsequently, the samples were rinsed with ethanol to remove non-chemisorbed dye. Then the samples were soaked in an aqueous solution of 5 mL KOH (10 mM) for 30 min. The concentration of desorbed dye was measured spectroscopically (by using a Lambda XLS UV/VIS spectrophotometer, PerkinElmer) at $\lambda$ = 502 nm.

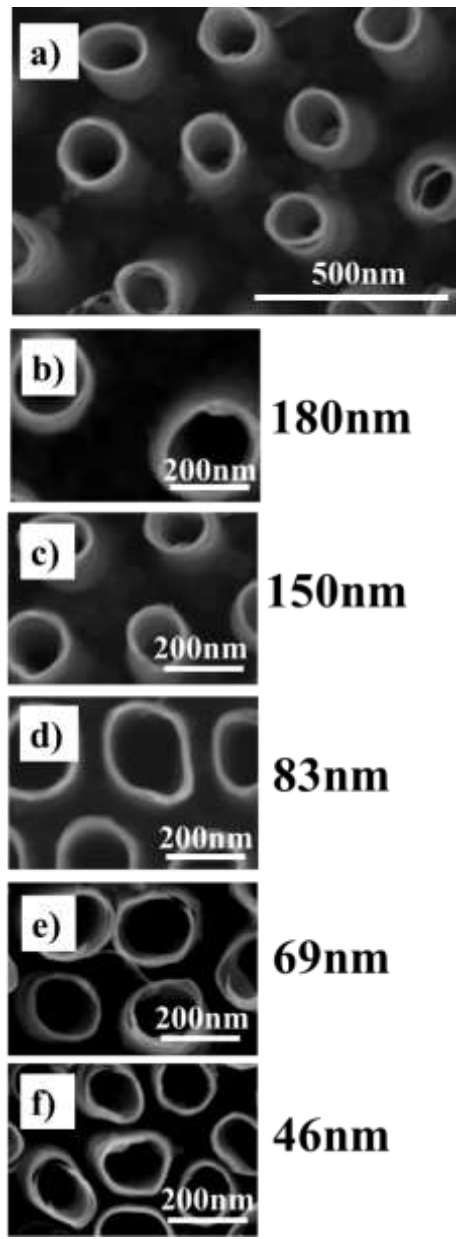

**Figure S1**. SEM images of different intertube distances formed by anodization of titanium in diethylene glycol based electrolyte.

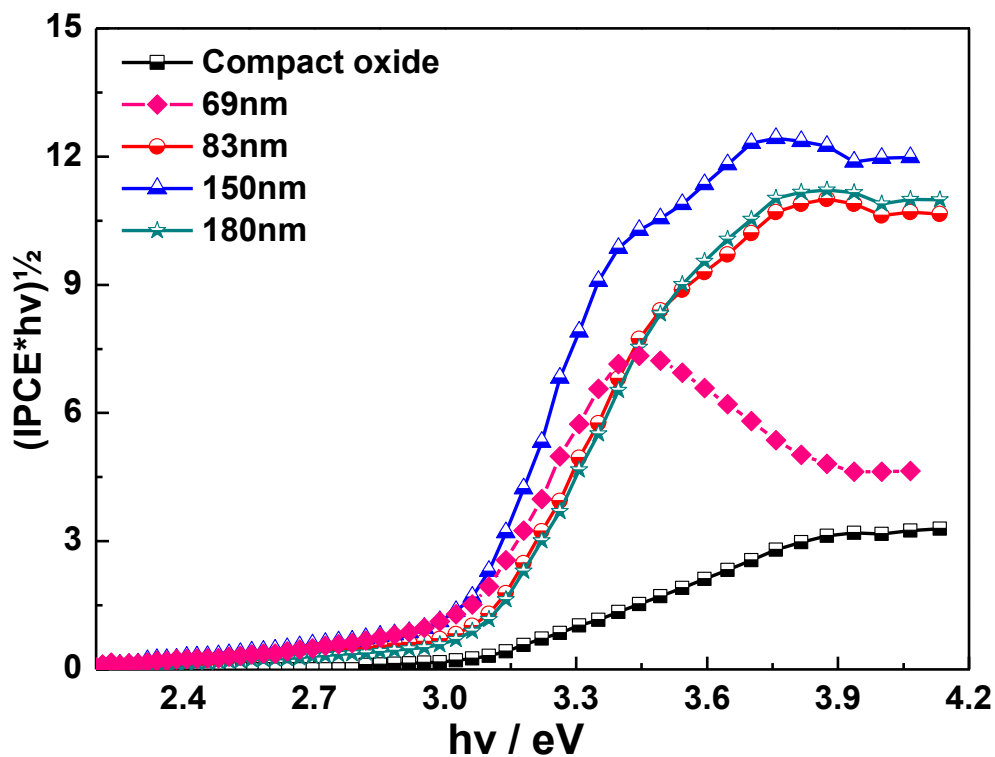

**Figure S2.** Bandgap evaluation determined from IPCE measurements, for compact oxide and spaced TiO$_2$ NTs with different intertube distances.

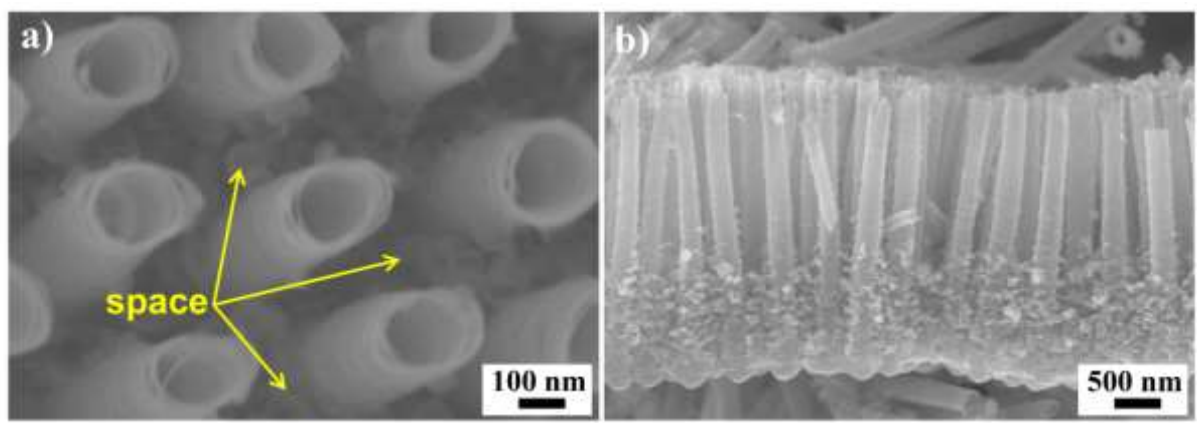

**Figure S3.** SEM images of a) top and b) cross-section of spaced TiO$_2$ NTs. The tube growth mechanism here is that spacing results from the initial TiO$_2$ NTs growth, where a selection mechanism similar to small nanotubes suppression in the case of conventional TiO$_2$ NTs layers occurs.

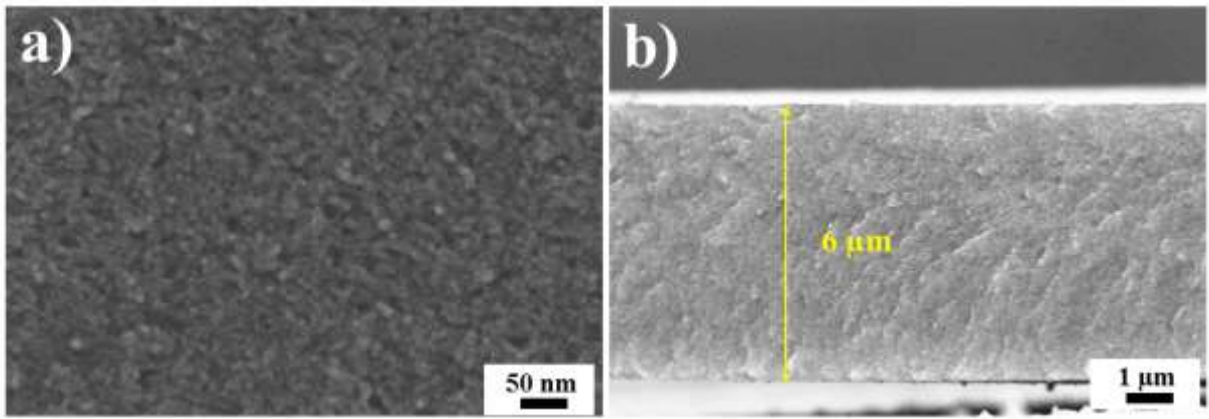

**Figure S4.** SEM images of a) top surface and b) cross-sectional of $TiO_2$ nanoparticle layers prepared by doctor blading method on fluorine-doped tin oxide (FTO).

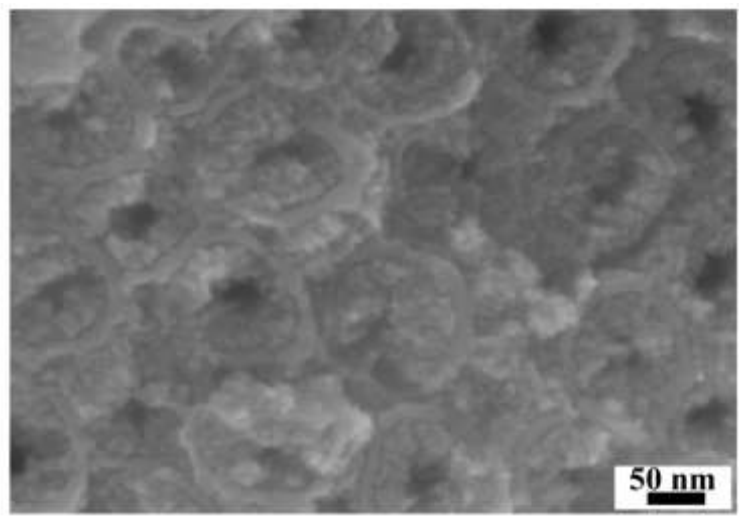

**Figure S5** SEM image of reference $TiO_2$ NTs decorated with 2 layers of nanoparticles: already after 2 times completely filled in a not fully defined structure. Photocatalytic $H_2$ generation of this sample showed only half of that measure for spaced NTs (with 2 times decoration of nanoparticles).

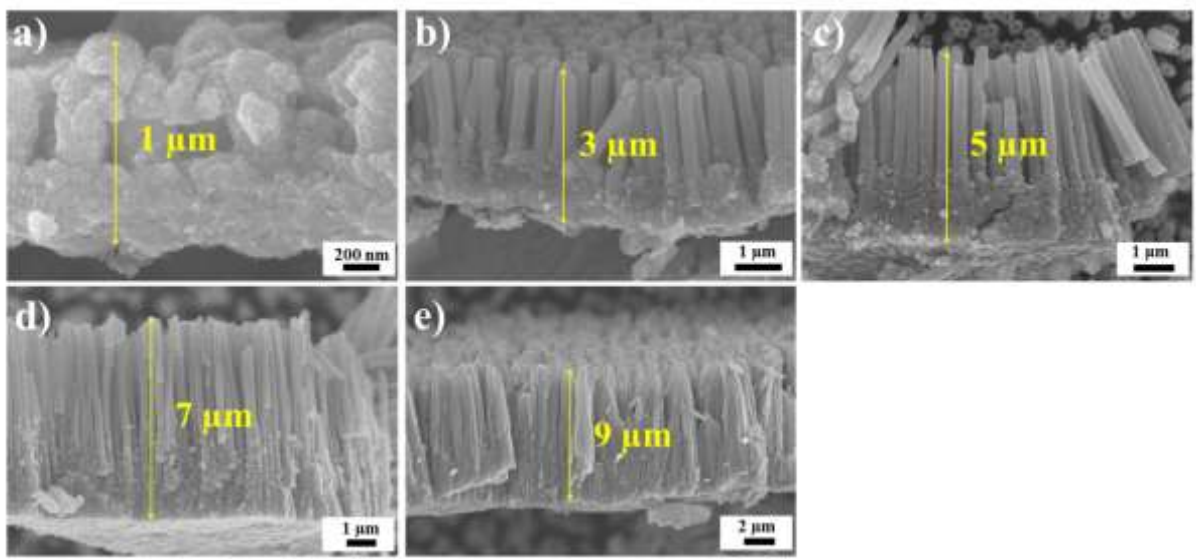

**Figure S6.** Different thicknesses of spaced $TiO_2$ NTs decorated with 3 layers of $TiO_2$ nanoparticles and 1 nm-thick Pt layers: a) 1, b) 3, c) 5, d) 7 and e) 9 μm.

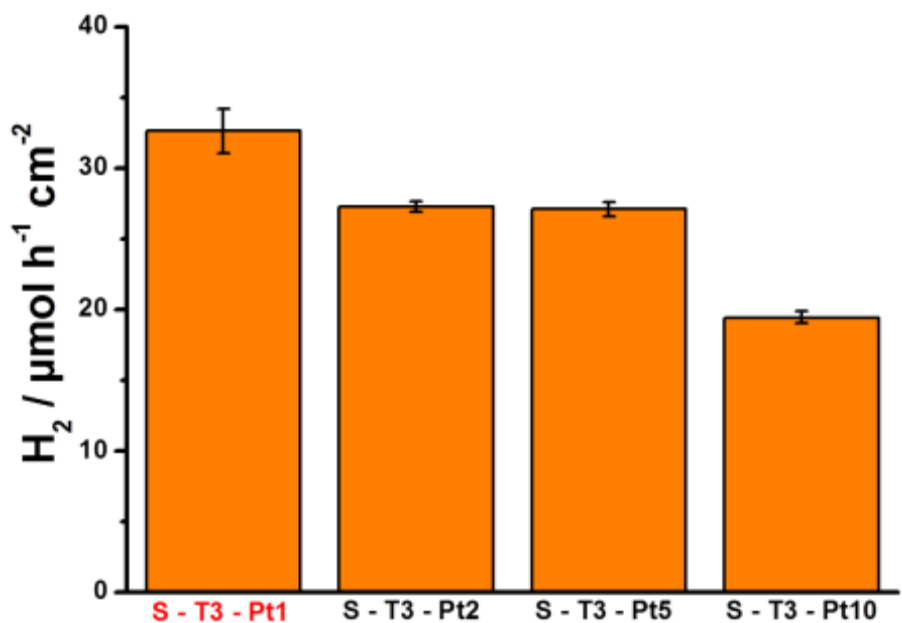

**Figure S7** Photocatalytic $H_2$ evolution measured for spaced $TiO_2$ NTs after decoration with 3 layers of $TiO_2$ nanoparticles and different amounts of Pt. Clearly, 1-nm thick Pt layer led to a maximized photocatalytic $H_2$ generation. Larger amounts of Pt induced larger shading effect of $TiO_2$; this means the light absorption of $TiO_2$ was hindered because Pt covered the oxide surface.

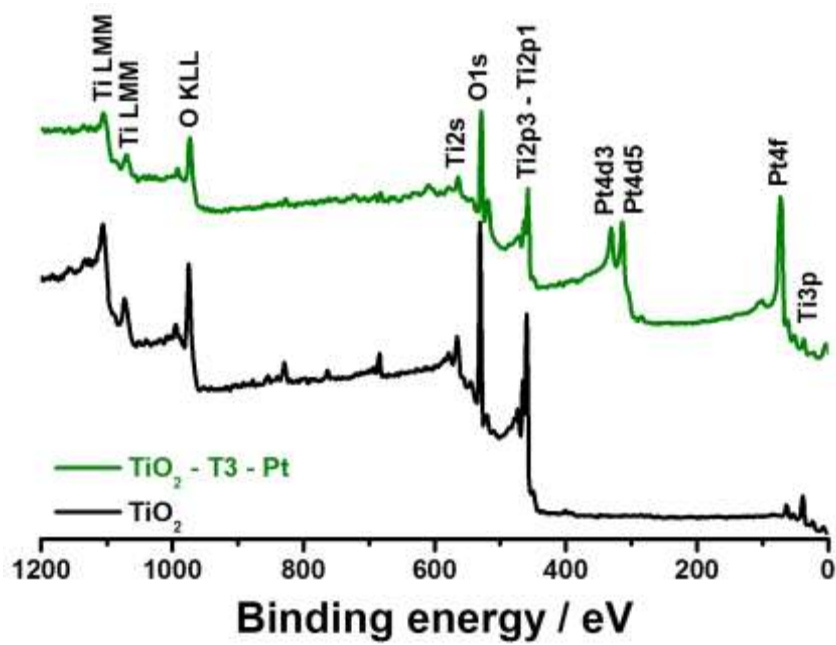

**Figure S8.** XPS survey of $TiO_2$ NTs, and $TiO_2$ NTs decorated with $TiO_2$ nanoparticles and 10 nm-thick Pt layer. The successful deposition of Pt is confirmed by the appearance of signals peaking at 72, 315, and 332 eV in the XPS spectra, which correspond to the binding energies of Pt 4f, Pt 4d5 and 4d3, respectively.

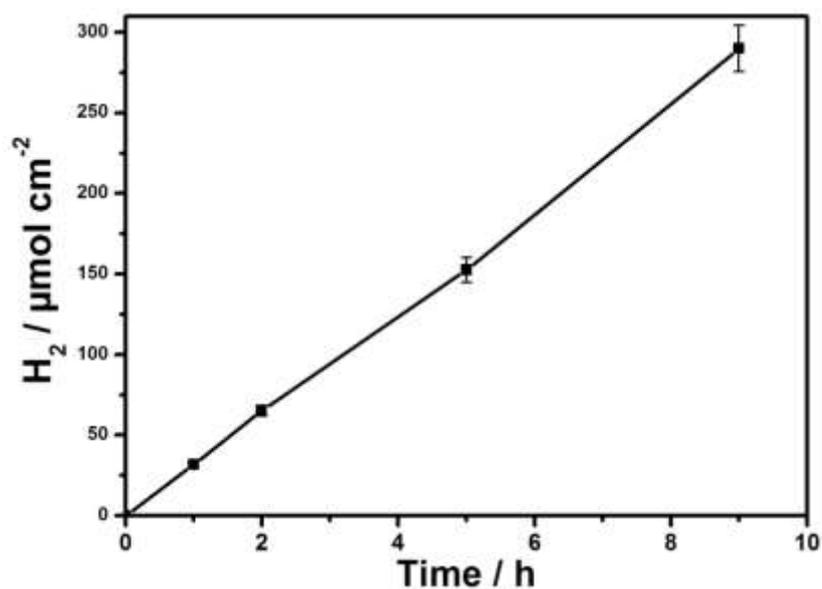

**Figure S9.** Photocatalytic $H_2$ evolution rate measured for spaced $TiO_2$ NTs after decoration with 3 layers of nanoparticles and 1 nm thick Pt.